# The Artificial Intelligence Act: critical overview

Nuno Sousa e Silva[*]

**Abstract**:
This article provides a critical overview of the recently approved Artificial Intelligence Act. It starts by presenting the main structure, objectives, and approach of Regulation (EU) 2024/1689. A definition of key concepts follows, and then the material and territorial scope, as well as the timing of application, are analyzed. Although the Regulation does not explicitly set out principles, the main ideas of fairness, accountability, transparency, and equity in AI underly a set of rules of the regulation. This is discussed before looking at the ill-defined set of forbidden AI practices (manipulation and e exploitation of vulnerabilities, social scoring, biometric identification and classification, and predictive policing). It is highlighted that those rules deal with behaviors rather than AI systems. The qualification and regulation of high-risk AI systems are tackled, alongside the obligation of transparency for certain systems, the regulation of general-purpose models, and the rules on certification, supervision, and sanctions. The text concludes that even if the overall framework can be deemed adequate and balanced, the approach is so complex that it risks defeating its own purpose of promoting responsible innovation within the European Union and beyond its borders.

---

[*] Lawyer and Assistant Professor at the Portuguese Catholic University (Porto). E: nsilva@ucp.pt W: www.nss.pt.

### A. Introduction

The rapid technological evolution of recent decades – generating a vast collection of digitized and accessible information (made possible by the Internet) and advances in terms of hardware and software – has allowed certain mathematical techniques (so-called machine learning) to become revolutionary. This is at the root of the dizzying developments in Artificial Intelligence that have taken place in the last few years.

However, despite the numerous advantages that this development brings,[1] the catastrophist tone has gained prominence.[2]

In the second decade of the 21st century, safety in Artificial Intelligence (hereinafter "**AI**") has established itself as an interdisciplinary branch of study, going beyond ethical considerations.[3] There are discussions regarding the transparency and explainability of decisions made by AI systems,[4] the potential for discrimination or injustice in the use of these systems,[5] and the challenges to control and align AI systems with human values.[6] There is a pressing need to guarantee the robustness and technical quality of AI.[7] The extractive practices of both data (some of it protected by intellectual

---

[1] Among many others, the acceleration of drug development (J. Jumper et al., 'Highly accurate protein structure prediction with AlphaFold' *Nature* 596 (2021) pp. 583-589) and vaccines (A Sharma, et al. *Artificial Intelligence-Based Data-Driven Strategy to Accelerate Research, Development, and Clinical Trials of COVID Vaccine.* BioMed research international (2022)), the fight against climate change (J. Cowls, et al. 'The AI gambit: leveraging artificial intelligence to combat climate change–opportunities, challenges, and recommendations' in AI & Society 38 (2023) pp. 283-307) and the creation of new materials (Phil De Luna (ed.), *Accelerated Materials Discovery: How to Use Artificial Intelligence to Speed Up Development* (De Gruyter 2022)).

[2] Among the most influential works along these lines are Nick Bostrom, *Superintelligence: Paths, Dangers, Strategies* (OUP 2014) and, earlier, Ray Kurzweil, *The Singularity Is Near: When Humans Transcend Biology* (Viking 2005). For a more balanced view, see Henry A Kissinger/Eric Schmidt/Daniel Huttenlocher, *The Age of AI: And Our Human Future* (Little, Brown and Company 2021).

[3] R.V. Yampolskiy, 'Artificial intelligence safety engineering: Why machine ethics is a wrong approach' in AAVV, *Philosophy and Theory of Artificial Intelligence* (Springer 2013) pp. 389-396.

[4] This is what is known as XAI (*explainable* AI). On the wider topic cf. Frank Pasquale, *The Black Box Society: The Secret Algorithms That Control Money and Information* (Harvard University Press 2016). Discussing the existence of a right to explanation under Article 22 GDPR, see the debate between Sandra Wachter / Brent Mittelstadt / Luciano Floridi, 'Why a Right to Explanation of Automated Decision-Making Does Not Exist in the General Data Protection Regulation' International Data Privacy Law, vol. 7(2) (2017) pp. 76-99 and Gianclaudio Malgieri / Giovanni Comande, 'Why a Right to Legibility of Automated Decision-Making Exists in the General Data Protection Regulation' International Data Privacy Law, 2017, Vol. 7(4) pp. 243-265. The majority of author seem to agree that under the GDPR there is no right to a detailed explanation of the decision, but only to a statement of its basic criteria and parameters (AAVV, *General Data Protection: Article-by-article commentary* (Hart C. H. Beck 2023) p. 541).

[5] Among many, Cathy O'Neil, *Weapons of Math Destruction: How Big Data Increases Inequality and Threatens Democracy* (Brown 2016); Safiya Umoja Noble, *Algorithms of Oppression: How Search Engines Reinforce Racism* (NYU Press 2018); Meredith Broussard, *More than a Glitch: Confronting Race, Gender, and Ability Bias in Tech* (MIT Press 2023).

[6] See Brian Christian, *The Alignment Problem* (Atlantic Books 2020) and Stuart Russel, *Human Compatible: AI and the Problem of Control* (Penguin 2019).

[7] Max Tegmark, *Life 3.0: Being Human in the Age of Artificial Intelligence* (Penguin 2017).

property rights) and minerals and the energy consumption of AI are also a matter of concern.[8]

In recent years, lawyers and politicians have started to consider laws to deal with the multiple challenges of AI. The issues are manyfold and have a subatantial impact on fundamental rights (freedom, work and employment, privacy, equality and non-discrimination, democratic participation, access to justice, freedom of expression and information, political organization, environmental protection), civil and criminal liability, personal data protection, privacy and personality rights, intellectual property, competition law, environmental law, criminal law, tax law and administrative law.[9]

Although regulatory initiatives are taking place all over the world, the European Union has taken the lead.[10] On February 16, 2017, the European Parliament adopted a resolution with recommendations to the European Commission on civil law rules on robotics.[11] This resolution recognizes the dangers and opportunities of robotics and artificial intelligence and makes various suggestions for their regulation, urging the Commission to present a legislative proposal on legal issues related to the development and use of robotics and Artificial Intelligence. Annexed to this document were recommendations on the content of such a proposal – including the definition of a robot, the creation of a registration system managed by a European agency, rules on civil liability, insurance and guarantee funds and the establishment of interoperability rules – and a "Robotics Charter", a voluntary code of conduct aimed at robotics researchers and *designers*. This 2017 resolution accelerated the discussion on legal issues related to artificial intelligence and robotics.[12]

---

[8] KATE CRAWFORD *Atlas of AI: Power, Politics, and the Planetary Costs of Artificial* Intelligence (Yale University Press 2021).

[9] Books on the Law and/of AI have multiplied. Initially, the study (and the European Parliament's approach) focused mainly on robotics, and the general works include UGO PAGALLO, *The Laws of Robots: Crimes, Contracts, and Torts* (Springer 2013); ALAIN BENSOUSSAN/JÉRÉMY BENSOUSSAN, *Droit des Robots* (Larcier 2015) and RYAN CALO/MICHAEL Froumkin/ IAN KERR (EDS), *Robot Law* (EE 2016). In fact, the tendency to focus analysis on robotics extended beyond law, as evidenced by PATRICK LIN/KEITH ABNEY/GEORGE A. BEKEY (eds), *Robot Ethics: The Ethical and Social Implications of Robotics* (MIT Press 2011). These books mainly dealt with personality, crime, contracts and torts (liability). Others, such as MOISÉS BARRIO ANDRÉS (eds), *Derecho de los Robots* (Wolters Kluwer 2018), have gone further, also dealing with issues of employment law, financial and tax law, health law and its impact on the legal professions. Still under a perspective of Law and Robotics, but focusing on Artificial Intelligence, cf. JACOB TURNER, *Robot Rules* (Palgrave 2019) and RYAN ABBOTT, *The Reasonable Robot* (Cambridge University Press 2020). In line with the more general trend, authors have come to prefer AI-centered analysis. More general books include MATT HERVEY/MATTHEW LAVY (eds.), *The Law of Artificial Intelligence* (Sweet & Maxwell 2020); WOODROW BARFIELD / UGO PAGALLO, *Advanced Introduction to Law and Artificial Intelligence* (Edward Elgar 2020); WOODROW BARFIELD / UGO PAGALLO (eds), *Research Handbook on the Law of Artificial Intelligence* (Edward Elgar 2020); JAN DE BRUYNE / CEDRIC VANLEENHOVE (eds.), *Artificial Intelligence and the Law* (Intersentia 2021); HOEREN / PINELLI, *Künstliche Intelligenz – Ethik und Recht* (C. H. Beck 2022); and CHARLES KERRIGAN, *Artificial Intelligence: Law and Regulation* (Edward Elgar 2022). EBERS/HEINZE/KRÜGEL/STEINRÖTTER, *Künstliche Intelligenz und Robotik* (C.H. Beck 2020) is noteworthy for its breadth and depth, with over a thousand pages of sectoral analysis. There are also empirical studies, critical theories and law and economics (e.g. GEORGIOS ZEKOS, *Economics and Law of Artificial Intelligence* (Springer 2021)).

[10] Beyond the EU, on 17 May 2024 the Framework Convention on Artificial Intelligence and Human Rights, Democracy and the Rule of Law was approved by the Council of Europe ("**CoE Convention**"). In the same month, the 2019 OECD guidelines (Recommendation on Artificial Intelligence) were revised (C/MIN(2024)16/FINAL). In the US, there is sectoral legislation, initiatives (e.g. USC 15 Chpater 19 – National Intelligence Initiative), state legislation and executive orders, but no general federal law has yet been passed. Some countries, such as Australia, Japan, Israel, Singapore and India, have followed *soft law* approaches, complemented by sectoral interventions. There have been some proposals for legislation, for example in Brazil and Canada. In July 2023, Peru adopted Law 31814 to promote the use of AI. For a follow-up on legislative and regulatory developments in this area, see. https://www.whitecase.com/insight-our-thinking/ai-watch-global-regulatory-tracker

[11] (2015/2103(INL)).

In the following year, the Commission presented two communications "Artificial Intelligence for Europe" [13] and "Coordinated Plan for Artificial Intelligence" [14]. Resolutions, studies and reports followed and the "White Paper on Artificial Intelligence" presented by the Commission in February 2020 set the approach for the upcoming proposals.[15]

On October 20, 2020, the European Parliament adopted a resolution with recommendations to the Commission on the civil liability regime applicable to artificial intelligence.[16] This document contained the text of a draft regulation on liability for the operation of AI systems.[17] On September 28, 2022, the European Commission presented two proposals: a revision of the Product Liability Directive, which aims to replace Directive 85/374/EC[18] and a new Directive on the adaptation of non-contractual civil liability rules to artificial intelligence.[19] These are still under discussion.

However, the main regulatory approach to this phenomenon is the Artificial Intelligence Regulation, known as the **AI Act.**[20] This regulation stems from a proposal presented by the European Commission in April 2021.[21] The proposal was the subject of intense negotiations (including a 36-hour marathon between representatives of the European Commission, European Parliament and Council), far-reaching amendments and a *corrigendum* (of April 19, 2024), was approved on June 13, 2024, and published on July 12 under the number 2024/1689.[22]

---

[12] On the state of the subject at that time, see the text (in portuguese) Nuno Sousa e Silva, 'Direito e Robótica: Uma primeira *aproximação*' Revista da Ordem dos Advogados [2017] pp. 485-551.
[13] COM(2018)237 final of April 25, 2018.
[14] COM(2018)795 final, of December 7, 2018.
[15] With the subtitle "A European approach to excellence and trust" (COM(2020)65 final)
[16] 2020/2014(INL).
[17] Recital 9 of this proposal reads: "*Council Directive 85/374/EEC ("Product Liability Directive") has proven for more than 30 years to be an effective means of obtaining compensation for damage caused by a defective product. It should therefore also be used with regard to civil liability actions by a party suffering loss or damage against the producer of a defective AI system.*" In fact, this Directive does not apply well to *software* and, more generally, to digital content or goods with digital content (including artificial intelligence agents and autonomous robots). In addition, there are some restrictions and practical obstacles to obtaining compensation.
[18] COM(2022)495 final.
[19] COM(2022)496 final.
[20] This text is referred to as the "**Regulation**" and to which, unless otherwise indicated or contextualized, the rules quoted without further indication belong. The legislative basis used is twofold: Articles 16 (on data protection) and 114 (on the internal market), both of the TFEU.
[21] COM(2021)206 final. For a description of the background and main features of the evolution of the proposals up to 2023, see Nikos Th. Nikolinakos, *EU Policy and Legal Framework for Artificial Intelligence, Robotics and Related Technologies–The AI Act* (Springer 2023) and Carmen Muñoz García, *Regulación de la inteligencia artificial en Europa* (Tirant lo Blanch 2023).

This article aims to provide a critical overview of the main aspects of this Regulation.

**B. Structure, objectives and approach**

The Regulation is an example of the so-called "regulatory brutality" trend.[23] As will become clear, this piece of legislation is particularly complex, involving 68 definitions, 113 articles, 13 annexes and 180 recitals. The penalties are severe (up to 7% of the offender's global revenue or 35 million euros), the territorial scope of application is particularly broad, and supervision is carried out at national and EU level, establishing a new regulatory architecture, which includes the *EU AI Office*, the EU AI *Board*, an advisory forum and a scientific panel of independent experts (arts. 64 ff.) and, at national level, at least one national notifying authority and one national market surveillance authority (Art. 70).

This legislative instrument is made up of 13 chapters: 1) general provisions; 2) prohibited practices; 3) high-risk systems; 4) transparency obligations for certain types of systems; 5) general purpose models; 6) measures in support of innovation; 7) governance; 8) high-risk system database; 9) post-market monitoring, information sharing, and market surveillance; 10) codes of conduct and guidelines; 11) delegation of powers and Committee procedure; 12) sanctions and 13) final provisions.

The major division of the Regulation is based on a risk classification of AI systems.[24] This classification considers the uses or applications of AI systems. It is, therefore, a question of knowing what the system is designed for, the so-called "intended purpose," defined in Art. 3/12 as *"the use for which an AI system is intended by the provider, including the specific context and conditions of use, as specified in the information supplied by the provider in the instructions for use, promotional or sales materials and statements, as well as in the technical documentation"*. Thus, the same algorithms and software applied both in system A and system B can lead to a different risk classification.[25] The approach is not on the technology but rather on the goal of each system. Conversely, the provider can exclude the application of certain rules or even the Regulation as a whole if it is careful and explicit in the instructions and materials it makes available.[26]

---

[22] Regulation (EU) 2024/1689 of the European Parliament and of the Council of 13 June 2024 creating harmonized rules on artificial intelligence and amending Regulations (EC) No 300/2008, (EU) No 167/2013, (EU) No 168/2013, (EU) 2018/858, (EU) 2018/1139 and (EU) 2019/2144 and Directives 2014/90/EU, (EU) 2016/797 and (EU) 2020/1828 (Artificial Intelligence Regulation).

[23] V. PAPAKONSTANTINOU/PAUL DE HERT, 'The Regulation of Digital Technologies in the EU: The law-making phenomena of "act-ification", "GDPR mimesis" and "EU law brutality"' Technology and Regulation [2022] pp. 48-60.

[24] "Risk" is defined in Article 3/2 of the Regulation as " *'the combination of the probability of an occurrence of harm and the severity of that harm;*". On the risk-based regulatory approach *see* GIOVANNI DE GREGORIO / PIETRO DUNN, 'The European risk-based approaches: Connecting constitutional dots in the digital age' Common Market Law Review vol. 59(2) (2022) pp. 473-500. Criticizing the notion of risk in the context of the regulation see MARCO ALMADA / NICOLAS PETIT, 'The EU AI act: a medley of product safety and fundamental rights?' RSC Working Paper 2023/59 pp. 19-20.

There are essentially two levels of risk: intolerable risk (which leads to the prohibition of certain practices or uses of AI systems - Article 5)[27] and high-risk.[28] Most of the rules are aimed at high-risk AI systems. As we shall see, Article 5 presents difficulties of interpretation and delimitation. It is therefore essential to look at Article 6, which defines high-risk systems, to understand the scope of the prohibited practices. If the Regulation considers a certain use of the AI system to be high-risk, then it cannot be included in the prohibited practices. In other words, article 6 is particularly important to define the scope of article 5.

The Regulation also regulates so-called general purpose AI models, i.e. "*an AI model (...) that displays significant generality and is capable of competently performing a wide range of distinct tasks regardless of the way the model is placed on the market and that can be integrated into a variety of downstream systems or applications, except*

---

[25] There will often be difficulty in determining what is the use in question - if the system has several possible applications and the Regulation applies to the entire value chain, could that system have different levels of risk along the chain? The answer must be yes. As noted, what matters for the classification is the intended use. When the system was designed for a given, low-risk use is actually being used for a high-risk application, Article 25 provides that this change of purpose can change the qualification of the person who made it, changing from "deployer" (the user) to "provider" (the person primarily responsible for ensuring compliance with the Regulation). In addition, the Regulation deals with general purpose models (Article 51 ff.), which can be used for many different purposes.

[26] Article 8. Even so, the Regulation obliges the producer of a high-risk system to have a risk management system, which includes (Art. 9/2/b)) the estimation and assessment of the risks that may arise from "reasonably foreseeable misuse", defined as " *use of an AI system in a way that is not in accordance with its intended purpose, but which may result from reasonably foreseeable human behaviour or interaction with other systems, including other AI systems*" (Art. 3/13).

[27] One might wonder if this approach makes sense. If the same application or practice took place without the use of AI systems, would it be legal? If the answer is no, then the association with AI systems is irrelevant. In fact, I submit that Article 5 is about regulating conducts and would not need to be AI-specific

[28] Article 50 does not refer to "low risk" or "limited risk", it applies in light of the use in question, regardless of the risk classification of the system. It is often pointed out that there are AI systems, such as video games and *spam* filters, which are not covered by the Regulation and would constitute another category of "no risk". think it would be better to just point out that t(hese systems are not covered by the Regulation. Nevertheless, recital 27 hints at voluntary compliance. MARCO ALMADA / NICOLAS PETIT, (n 24), pp. 8-9 mention three tiers: intolerable risk (art. 5), high-risk (covered by the Regulation) and other AI systems (which are not covered by the Regulation, but are subject i.a. to Regulation 2023/988 on general product safety).

*AI models that are used for research, development or prototyping activities before they are placed on the market*" (art. 3/63), in particular those that present a systemic risk (arts. 51 ff.).

The approach taken in the Regulation is in line with legislation on product safety,[29] namely Regulation (EU) 2023/988 of 10 May 2023 on general product safety,[30] and sectoral regulatory instruments on toys,[31] cosmetics,[32] and medical devices.[33] AI systems are regarded as goods, and high-risk systems must bear a conformity mark (*CE* – short for *conformité européenne*) which confirms that there has been a verification and that the (high-risk) AI system complies with the applicable EU legislation (art. 48).[34] The simplest way to avoid ambiguities and interpretative difficulties will be to follow the standards and technical norms approved under the Standards Regulation,[35] thereby benefiting from a presumption of conformity (arts. 40/1 and 42/2).[36]

Nevertheless, the Regulation takes into account the complexity (and sophistication) of Artificial Intelligence. To use Laura Caroli's words, "[an AI system] is not a toaster". This is why the AI Act presents considerable deviations from classic product safety laws, namely by imposing duties on users of the systems (art. 26) and, in some cases, requiring an impact assessment on fundamental rights (art. 27). The

---

[29] Michael Veale / Frederik Zuiderveen Borgesius, 'Demystifying the Draft EU Artificial Intelligence Act-Analysing the good, the bad, and the unclear elements of the proposed approach' Computer Law Review International (2021) p. 98. In this sense, the AI Act makes copious references to Regulation (EU) 2019/1020 on market surveillance and product conformity.

[30] In 2008 the EU adopted the so-called "New Legislative Framework", an updated legislative package of general rules for ensuring product safety and conformity, accompanied by special rules for certain categories (to date 26 categories, including elevators, construction material, explosives, radio, fertilizers, batteries, machinery and *drones*). The Regulation is now part of this category of legislation.

[31] Directive 2009/48/EC of the European Parliament and of the Council of June 18, 2009 on the safety of toys

[32] Regulation (EC) No 1223/2009 of the European Parliament and of the Council of 30 November 2009 on cosmetic products.

[33] Regulation (EU) 2017/745 of the European Parliament and of the Council of April 5, 2017 on medical devices. As stated in recital 19 of this regulation, "*It is necessary to clarify that software in its own right, when specifically intended by the manufacturer to be used for one or more of the medical purposes set out in the definition of a medical device, qualifies as a medical device, while software for general purposes, even when used in a healthcare setting, or software intended for life-style and well-being purposes is not a medical device. The qualification of software, either as a device or an accessory, is independent of the software's location or the type of interconnection between the software and a device.*" Peter Feldschreiber (ed.), *The Law and Regulation of Medicines and Medical Devices* (OUP 2021).

[34] The CE marking rules are contained in Regulation (EC) No 765/2008 of the European Parliament and of the Council of 9 July 2008 setting out the requirements for accreditation and market surveillance relating to the marketing of products. The rules on standards are contained in Regulation (EU) No 1025/2012 of the European Parliament and of the Council of 25 October 2012 on European standardization.

[35] Regulation (EU) No 1025/2012 of the European Parliament and of the Council of 25 October 2012 on European standardization, amending Council Directives 89/686/EEC and 93/15/EEC and Directives 94/9/EC, 94/25/EC, 95/16/EC, 97/23/EC, 98/34/EC, 2004/22/EC, 2007/23/EC, 2009/23/EC and 2009/105/EC of the European Parliament and of the Council and repealing Council Decision 87/95/EEC and Decision 1673/2006/EC of the European Parliament and of the Council.

[36] Michael Veale / Frederik Zuiderveen Borgesius, (n 29), p. 105 point out that this will probably be the path followed by most producers.

Regulation is therefore a hybrid, combining an approach typical of rights-legislation such as the GDPR with another, typical of regulatory law. However, with the exception of the right to lodge a complaint (art. 85) and an explanation of the role of the AI system in certain decisions (art. 86), this Regulation does not establish subjective rights.

Although this is an "Artificial Intelligence" regulation, it seems to me that many of these practices and actions, especially the prohibited ones, would already be covered by the existing regulatory framework, namely the Digital Services Regulation,[37] the General Data Protection Regulation,[38] the rules of fair competition and consumer protection, including Advertising Law and, more generally, the rules protecting personality rights and Fundamental Rights.[39] We must not forget that the regulation of Artificial Intelligence does not begin or end with this Regulation, despite its undeniable importance.[40]

**C. Concepts**

The Regulation has taken a maximalist approach to definitions, defining terms that are already part of the European acquis such as "personal data", "non-personal data", "profiling", "biometric data", enshrining unhelpful definitions such as "AI literacy" and terms that are self-explanatory such as "publicly accessible space", "training data" or "instructions for use".[41]

On the other hand, the concept "law enforcement" is important but not obvious This term, which appears 98 times in the Regulation, is defined in Article 3/46 as "*activities carried out by law enforcement authorities or on their behalf for the prevention, investigation, detection or prosecution of criminal offences or the execution of criminal penalties, including safeguarding against and preventing threats to public security*", with "law enforcement authority" being "any public authority competent for the prevention, investigation, detection or prosecution of criminal offences or the execution of criminal penalties, including the safeguarding against and the prevention of threats to public security; or any other body or entity entrusted by Member State law to exercise public authority and public powers for the purposes of the prevention, investigation, detection or prosecution of criminal offences or the execution of criminal penalties, including the safeguarding against and the prevention of threats to public security" (art. 3/45). In other words, when the Regulation refers to law enforcement, it is essentially referring to police activity.

---

[37] Regulation (EU) 2022/2065 of the European Parliament and of the Council of October 19, 2022 on a single market for digital services and amending Directive 2000/31/EC (Digital Services Regulation).
[38] Regulation (EU) 2016/679 of the European Parliament and of the Council of April 27, 2016 on the protection of natural persons with regard to the processing of personal data and on the free movement of such data, and repealing Directive 95/46/EC (General Data Protection Regulation).
[39] Cfr. STEFAN SCHEURER, 'Artificial Intelligence and Unfair Competition - Unveiling an Underestimated Building Block of the AI Regulation Landscape' GRUR Int vol. 70(9) (2021) pp. 834-845.
[40] Even in terms of product safety in the internal market, recital 166 points out that "*it is important that AI systems related to products that are not high-risk in accordance with this Regulation and thus are not required to comply with the requirements set out for high-risk AI systems are nevertheless safe when placed on the market or put into service. To contribute to this objective, Regulation (EU) 2023/988 of the European Parliament and of the Council (53) would apply as a safety net*".
[41] See, respectively, Article 3(50), (51), (52), (34), (56), (44), (29), and (15). On the other hand, "widespread infringement" (art. 3/61) and "deep fakes" (arts. 3/60) are defined, but these terms are used only once in the Regulation (respectively arts. 73/3 and 50/4).

For a proper understanding of the regulation, it is necessary to understand the definition of an Artificial Intelligence system and analyze the various categories of subjects.

**I. Artificial Intelligence System**

The first challenge for regulation was to find a suitable definition of Artificial Intelligence. Many definitions associate intelligence with human intelligence, the ability to use reasoning to achieve goals. Other perspectives approach the concept through the programming techniques used.[42] After much discussion, the Regulation ended up adopting the definition of "Artificial Intelligence systems", which replicates the updated OECD definition: " *a machine-based system that is designed to operate with varying levels of autonomy and that may exhibit adaptiveness after deployment, and that, for explicit or implicit objectives, infers, from the input it receives, how to generate outputs such as predictions, content, recommendations, or decisions that can influence physical or virtual environments*" (art. 3/1).[43]

---

[42] This was the much-criticized original approach of the European Commission.
[43] The concept also corresponds to that used in Article 2 of the CoE Convention. On updating the OECD definition, see the *Explanatory Memorandum on The Updated OECD Definition of an AI System* (March 2024), which prefers to use the notion of AI *systems* for regulatory purposes. This perspective is in line with the US Executive Order on Artificial Intelligence (*Executive Order 14110 on Safe, Secure, and Trustworthy Development and Use of Artificial Intelligence)*, but has some notable differences. The Presidential Order, which generalizes the approach of Executive Order 13960 (which was directed only at federal agencies), is based mainly on cybersecurity requirements, monitoring and technical quality of systems and defines Artificial Intelligence. In section 3 b) of EO 14110 as "a *machine-based system that can, for a given set of human-defined objectives, make predictions, recommendations, or decisions influencing real or virtual environments. Artificial intelligence systems use machine- and human-based inputs to perceive real and virtual environments; abstract such perceptions into models through analysis in an automated manner; and use model inference to formulate options for information or action*". As you can see, the US Executive Order refers to the human definition of objectives, which is not required in the OECD and AI Act definitions. Luciano Floridi, 'On the Brussels-Washington Consensus About the Legal Definition of Artificial Intelligence'. Philosophy & Technology (2023) vol. 36 (87) The definition of AI used in the ISO/IEC 22989:2022 (2022) standard is close: "*a technical and scientific field devoted to the engineered system that generates outputs such as content, forecasts, recommendations or decisions for a given set of defined objectives*".

This notion seems particularly broad and almost coincides with the concept of software. The distinction lies in the existence of some degree of **autonomy** and the mention to **inferences**. In this sense, recital 12 explains that "*the definition should be based on key characteristics of AI systems that distinguish it from simpler traditional software systems or programming approaches and should not cover systems that are based on the rules defined solely by natural persons to automatically execute operations*". To this end, it stresses that what is essential is the ability to make *inferences, i.e. the* possibility of processing or generating new data in contexts other than those in which the system was trained.[44] In other words, simple automations, formulas, static software or totally deterministic programming (*if x, then y*) are excluded.[45] As the notion is broad, in case of doubt the system analyzed should be considered an AI system.

It is important to stress that the regulation essentially concerns systems as a whole (including *hardware,* i.e. computers, sensors, peripherals and *software that does* not constitute artificial intelligence). **Systems** must be distinguished from **models**. As pointed out in recital 97: "*Although AI models are essential components of AI systems, they do not constitute AI systems on their own. AI models require the addition of further components, such as for example a user interface, to become AI systems. AI models are typically integrated into and form part of AI systems.*" While ChatGPT (from OpenAI) constitutes an AI system (including several layers of software, a graphical interface, servers, etc.), there are several models (which act as the system's "engine") that can integrate it (to date, and in the case of ChatGPT, three options are available: GPT 3.5, GPT 4 and GPT 4). It is possible to use one model and build systems with very different applications, purposes and modes of operation.[46]

**II. Subjects**

The Regulation mentions several roles that form part of the AI value chain: the provider, the importer, the distributor, the authorised representative, and the deployer, all of whom are covered by the generic notion of "**operator**". As we shall see, the Regulation applies to any provision of the system in the EU, even if it is free of charge.[47]

---

[44] Recital 12: "*... The capacity of an AI system to infer transcends basic data processing by enabling learning, reasoning or modelling*.". As highlighted in the *Explanatory Memorandum on The Updated OECD definition*, there are also inferences in the training phase, especially in the case of *unsupervised machine learning*.

[45] *Robotic Process Automation* (i.e. a way of automating repetitive processes, usually in a business context) will have to be analyzed specifically. In some cases, AI agents may be involved; in others, it is mere deterministic programming. In any case, it seems that most cases of RPA will not fall within the material scope of the Regulation as they are unlikely to present a relevant risk.

[46] It is above all to this extent that, as we shall see, the Regulation is also concerned with models. There is, however, a definition of "general-purpose AI system" in Art. 3/66 as "*an AI system which is based on a general-purpose AI model and which has the capability to serve a variety of purposes, both for direct use as well as for integration in other AI systems*". This concept is only used by the Regulation to refer to the modification of such a system to serve a (specific) purpose classified as high-risk (art. 25/1/c)).

The central subject, the main target of the AI Act rules, is the *provider*, defined in Article 3/3 as "*a natural or legal person, public authority, agency or other body that develops an AI system or a general-purpose AI model or that has an AI system or a general-purpose AI model developed and places it on the market or puts the AI system into service under its own name or trademark, whether for payment or free of charge;*". The central feature that defines someone as a provider is the fact that they offer an AI-system under their own name.[48] Providers, when not established in the EU, must fulfill their obligations through **authorised representatives** established in the EU (defined in art. 3/5), as provided for in arts. 22 (in the case of high-risk AI systems) and 54 (general purpose AI models).[49]

The user, except for those who use the system as part of a personal, **non-professional** activity,[50] is the "*deployer*" (art. 3/4) and also has obligations of their own, namely, to supervise the operation of the system (cf. arts. 26 and 50/3 and /4).

Importers, i.e. those persons located in the EU who place an AI system on the internal market (art. 3/6), will have certain obligations to verify and guarantee conformity, as well as to collaborate with the authorities (art. 23). The Regulation reserves the term "placing on the market" for the initial act making available of an AI system on EU territory (art. 3/9), with "making available" being defined as any supply in the context of a commercial activity (art. 3/10). Thus, importers carry out "placing on the market", while distributors (Art. 3/7) are engaged in "making available on the market" following importation.[51] Distributors are subject to obligations of verification and cooperation with the authorities that are very similar to those of importers (Art. 24).

---

[47] The Regulation refers to the territory of the Union and does not cover other countries in the European Economic Area.
[48] This will also include so-called OEMs (*Original Equipment Manufacturers*), who may not have had any role in the development of the system, but who integrate it into their product and/or present the AI system as their own. The qualification cannot be circumvented, however, by arguing that the "producer" of the system merely provides technical means. Of course, there will be dubious situations: when company A provides *middleware* to allow its customers to develop AI models, applications, or even systems and/or allows these models and applications to run on its infrastructure (servers), who is the provider? I think we can consider company A's *middleware* system as an AI system and company A as the provider of that system. However, the systems developed by each of company A's clients and eventually made available to third parties will eventually constitute separate systems of which company A's clients will be the providers. The situation can get complicated if company A provides a configurable/parameterizable AI system. In that case, considering Art. 25, whether those customers remain deployers or become providers of a new system will depend on the extent of the modifications made and/or the branding of that customer on the system. If these are signifcative, company A (provider of the original system) must "*closely cooperate with new providers and shall make available the necessary information and provide the reasonably expected technical access and other assistance that are required for the fulfilment of the obligations set out in this Regulation*" (art. 25/2). There is also provision for "mandatory contracting": according to Article 25/4 "*The provider of a high-risk AI system and the third party that supplies an AI system, tools, services, components, or processes that are used or integrated in a high-risk AI system shall, **by written agreement**, specify the necessary information, capabilities, technical access and other assistance based on the generally acknowledged state of the art, in order to enable the provider of the high-risk AI system to fully comply with the obligations set out in this Regulation".* Although recital 88 may give a different impression, I don't believe that Article 25/4 applies to those who merely provide models and I believe that the "mandatory" contracting provided for in this article should be interpreted restrictively (otherwise, absurdly, the supplier of cooling systems for the computers used to train an AI system or even the supplier of meals to *data scientists* could be covered).
[49] This obligation is similar to that laid down in Art. 27 GDPR.
[50] I anticipate that this exception will be interpreted restrictively. Thus, my use of an AI system to generate images for a conference presentation as a teacher or lawyer would not be covered.

Another concept, which is not defined but is included in the concept of operator, is that of "product manufacturer" (referred to in Article 2/1/e)). Given that what is at stake is the joint provision of a product and an AI system under one's own name or brand, product manufacturers should be considered providers.[52]

A person can become a provider if they "*put their name or trademark on a high-risk AI-system already placed on the market*" (art. 25/1/a)), "*make a substantial modification to a high-risk AI-system that has already been placed on the market or put into service, in such a way that it remains a high-risk AI-system*" (art. 25/1/b)) or "*modify the intended purpose (...) so that the AI-system concerned becomes a high-risk AI-system" (Art.* 25/1/c)).[53] Although the reverse is not expressly spelled out, changing the intended use of the AI system to one that is not considered high-risk will allow the modified system to escape the application of certain rules or even the Regulation as a whole.

**D. Scope of application**

Despite being a general regulation, the AI Act explicitly safeguards the application of the rest of the regulatory framework (art. 2, paragraphs 5, 7 and 9)[54] and allows complementary national rules to be adopted in certain areas, such as more favorable standards for the protection of workers (art. 2/11) or rules on the use of remote biometric identification systems (art. 5/5 and /10). In addition, the application of some legislation (art. 2/2, referring to the list in Section B of Annex I) and sectoral supervision (arts. 72 and 74) is reserved.[55] There are also matters that depend on implementing measures at national level, in particular the designation of national authorities and the supervisory framework (arts. 70 and 74), as well as the sanctions regime (art. 99/2). On the other hand, the Commission has a broad power to adopt delegated acts, complete and update the Regulation (arts. 7 and 97), and perform extensive evaluations and reviews (art. 112). The Commission will also draw up comprehensive guidelines on the Regulation (Art. 96) and encourage the development of codes of practice (Art. 56).

---

[51] This distinction will be more frequent when AI systems integrate hardware than with pure software. In any case, there are often software distribution agreements, including resale.
[52] In this sense, see art. 25/3.
[53] The notion of substantial modification is defined in Art. 3/23 as "*a change to an AI system after its placing on the market or putting into service which is not foreseen or planned in the initial conformity assessment carried out by the provider and as a result of which the compliance of the AI system with the requirements set out in Chapter III, Section 2 is affected or results in a modification to the intended purpose for which the AI system has been assessed*". This definition is close to the idea of purpose change set out in Article 6/4 of the GDPR. A system subject to a substantial modification is treated in the Regulation as a new system (cf. art. 43/4).
[54] In addition to these provisions, there are rules, such as art. 87, which expressly refers to other European legislation.

The AI Regulation is in line with the latest trend in digital single market regulation, having **extraterritorial** application.[56] According to Article 2/1, a minimum point of contact of the user or the result of the AI system with the territory of the Union is sufficient to trigger the applicability of the Regulation. Thus, if the result of an AI system is used in the EU or affects people located in the EU, this is enough for the Regulation to apply. On the other hand, the Regulation does not apply to anyone who develops AI systems in the EU, even for purposes prohibited by the Regulation, for use in third countries (i.e. there is no export control). Along the same lines, there is an obligation for providers established in third countries to appoint an authorised representative (Articles 22 and 54).

An important note in terms of **jurisdiction** concerns the decentralized nature of supervision. Except in the case of general-purpose AI models, which will be supervised by the European Commission, the competent national authorities will be responsible for dealing with all infringements that take place within their territory. Thus, the same provider and infringement may be subject to the concurrent jurisdiction of several national authorities.

Pursuant to article 2/6, research and development activities "in the laboratory" are excluded from the material scope of the Regulation (articles 57 ff. establish a complex set of rules for testing in a real environment). Activities prior to the system

---

[55] It should also be noted that the Regulation, in articles 102 to 110, amends various instruments of EU law.
[56] CHRISTOPHER KUNER, 'Protecting EU Data Outside EU Borders under the GDPR' Common Market Law Review 60 (2023) pp. 77-106. This approach by the European Union has contributed to the so-called "Brussels Effect", a term coined and described by ANU BRADFORD, *The Brussels Effect: How the European Union Rules the World* (OUP 2020). This expression alludes to the influential power of the European Union's regulatory acquis in matters such as competition law, environmental law, digital law and data protection. In these areas, the EU has been a pioneer in regulation and is often followed as a model in other jurisdictions. In addition, multinational companies end up adopting European rules as a global *compliance* standard. However, in the specific case of the Artificial Intelligence Regulation, it is far from clear whether the approach taken at EU level will have this effect (in the negative sense, see UGO PAGALLO, *Why the AI Act Won't Trigger a Brussels Effect* (2023) in https://ssrn.com/abstract=4696148).

being placed on the market or put into service are also not covered by the Regulation (art. 2/8).

The Regulation will also not apply to systems developed or used exclusively for military, national security or defence purposes (art. 2/3) or to use by public authorities of third countries and international organizations provided that these entities adequately safeguard fundamental rights (art. 2/4).

The subject of **open source** has been the subject of much debate.[57] "Domestic" uses are excluded, i.e. "*in the context of a personal activity of a non-professional nature*" (art. 2/10). However, making software (including the parameters of a model) available under open source licenses can also be done in a professional context.[58] The compromise solution is a limited exemption (Art. 2/12).[59] The key is the aforementioned difference between models and systems. The provision of open-source AI *models* enjoys certain exemptions under the Regulation. *Models* made available under open-source licenses are only required to comply with two obligations (copyright compliance policy and transparency regarding training data)[60] except in the case of general-purpose models with systemic risk (articles 25/4, 53/2 and 54/6). On the other hand, for AI *systems* covered by the Regulation (regardless of the level of risk), the fact that they are made available in *open source* is irrelevant. Simply put, the exemption is for models, not systems.

The application of these Regulations over time will be phased in. The Regulation entered into force on August 1, 2024, with the amendments to the legislation mentioned in articles 102 to 110 taking effect on that date. The general application of the Regulation is scheduled for August 2, 2026 (art. 113). There are, however, parts of the Regulation that will apply sooner. This is the case for the first two chapters (on prohibited practices), which will apply from February 2, 2025 (art. 113/a)), and the rules on the institutional framework, which will apply from August 2, 2025 (art. 113/b)).

---

[57] On the notion and history of *open source* see AMANDA BROCK (ed), *Open Source Law, Policy and Practice* (OUP 2022). As for AI, the debate is at various levels. Some advocate the need to restrict the circulation of information (and are proponents of what is known as *security through obscurity*), going so far as to compare the availability of code for certain systems to the availability of instructions for producing an atomic bomb. Others argue that openness is the most effective way of guaranteeing diversity, advancement, and even security. There is also considerable disagreement as to what is meant by *open source* in AI: whether it is enough to make the architecture and parameters of a model available (e.g. *open weights*) or whether the *dataset* used to develop it must also be made available. On the conceptual discussion in this area see ANDREAS LIESENFELD/MARK DINGEMANSE, 'Rethinking open source generative AI: open-washing and the EU AI Act' FAccT '24: Proceedings of the 2024 ACM Conference on Fairness, Accountability, and Transparency (June 2024) pp.1774-1787. In recitals 102 and 103, the Regulation seems to adopt a rather narrow notion of open source.
[58] In fact, in some contexts only companies with a lot of resources will be able to develop certain models (e.g. LLama developed by Meta).
[59] In fact, the text of art. 2/12 is completely useless: the exclusion provided for does not apply to the three types of systems covered by the Regulation.
[60] As explained in recital 104, the fact that a model is *open source* does not mean that one will have access to the training data or that respect for intellectual property rights has been guaranteed.

On the other hand, the rules on high-risk systems that are safety components of harmonized products (art. 6/1) will have a *vacatio legis of* 36 months and will only apply from August 2, 2027 (art. 113/c)). More importantly, the rules concerning high-risk AI systems will only apply to AI systems placed on the market after that date. AI systems already placed on the market, when they are considered high-risk, are exempt from the rules of the Regulation unless they undergo significant changes (Art. 111/2).[61] General purpose AI models placed on the market before August 2, 2025, will only be required to comply with the Regulation from August 2, 2027 (art. 111/3).[62]

**E. Principles**

Although not in the initial proposal, which was essentially aimed at determining prohibited practices and regulating high-risk applications, there was consideration of enshrining a set of general principles applicable to all operators and all AI systems subject to the Regulation.[63] In the final version, the only duty with such breadth is the obligation imposed on providers and implementers to ensure that people operating or using AI systems "*have a sufficient level of AI **literacy***" (art. 4).[64]

Nevertheless, those principles still underlie the requirements placed on high-risk systems (arts. 8 to 15) and their operators (arts. 16 to 27).

---

[61] In that sense, it is no longer the same system. It is unclear how the concept of "significant changes in their design" differs from "substantial modification" used in Articles 25 and 43/4. Recital 128 indicates that the concepts do not coincide. In any case, this rule, which gives a significant advantage to incumbent operators, is explained by the prohibition of retroactivity (what triggers the application of most of the Regulation's rules is the placing on the market). On the other hand, the prohibitions in Article 5, which refer to prohibited practices (and not system requirements) can and will be fully applicable to systems that are already on the market. In the case of certain "large-scale IT systems" of the European Union already in use, such as the Schengen IT system or the visa and travel information system (the list is in Annex X), which are already in operation, it is stipulated that they must be brought into conformity with the Regulation by December 31, 2030 (Article 111/1).

[62] On the other hand, models placed on the market after August 2, 2025 will have to comply with the rules "immediately" (art. 113/1/b)).

[63] In particular in Article 4a presented in May 2023 (COM(2021)0206 – C9 0146/2021 – 2021/0106(COD)), which set out the following principles: "a) human oversight and control; b) technical robustness and security; c) privacy and data governance; d) transparency; e) diversity, non-discrimination and *fairness*; f) social and environmental well-being". Articles 7 to 13 of the CoE Convention also set out the following principles: human dignity and autonomy, transparency and control, *accountability and* responsibility, equality and non-discrimination, protection of privacy and personal data, reliability and safe innovation. Many of these principles coincide with those listed in Article 5 of the GDPR, which will remain fully applicable whenever AI systems process personal data. Recital 27 of the AI Act mentions the "seven non-binding ethical principles" and encourages voluntary compliance with them.

[64] Article 20 of the CoE Convention also establishes a principle of promoting digital literacy.

At issue is a set of concerns developed in the interdisciplinary field known as AI safety or *FATE (Fairness, Accountability, Transparency, Ethics) AI*, including concerns of control, transparency, alignment, non-discrimination, robustness, and security.

Some principles are hard to parse. Of course, we are all in favor of *fairness*. The great difficulty, which is the field of philosophy and then politics, translating into the committed choice of each society at a certain time and place through positive law, lies in defining what is just, equitable, and fair. This problem is both conceptual and technical-mathematical.[65] In practical terms, not much can be drawn from this principle.

There are similar difficulties with algorithmic bias. Some of the known problems result from the poor quality of the data used (namely lack of representativeness or quantitative or qualitative insufficiency) or programming errors.[66] On the other hand, many problematic situations simply result from the system having been optimized to achieve a given beneficial or innocuous objective. For example, if an algorithm is designed to favor what an internet user pays more attention to, it could end up recommending alcoholic drinks (having indirectly detected that (s)he is an alcoholic), or promoting offensive or aggressive speech (since this is what most people will pay more attention to). These challenges, especially those posed by recommender systems, are already partially addressed in the Digital Services Act ("DSA").[67] In any case, the AI Act places significant emphasis on diversity and the prevention of discrimination and bias.[68] Putting an end to these occurrences is impossible, but there is an obligation to make adequate efforts to follow best practices to prevent easily avoidable mistakes.

---

[65] SORELLE A. FRIEDLER / CARLOS SCHEIDEGGER / SURESH VENKATASUBRAMANIAN, 'The (Im)possibility of fairness: different value systems require different mechanisms for fair decision making' Communications of the ACM. 64 (4) (2021) pp. 136-143.

[66] Examples abound, such as Google Photos' facial recognition system classifying black individuals as gorillas (in 2015), Amazon's recruitment tool prejudicing women (2018) and, more recently, in 2023, the iTutorGroup tool, used in recruitment, automatically rejecting applications from women over 55 and men over 60. The problem of algorithmic discrimination is widespread and reaches a large scale, as demonstrated by Z. OBERMEYER et al., 'Dissecting racial bias in an algorithm used to manage the health of populations' Science, (2019) 366(6464) pp. 447-453 on the health system in the USA. HILDE WEERTS et al, 'Algorithmic unfairness through the lens of EU non-discrimination law: Or why the law is not a decision tree'. Proceedings of the 2023 ACM Conference on Fairness, Accountability, and Transparency (2023) pp. 805-816 and PHILIPP HACKER, 'Teaching Fairness to Artificial Intelligence: Existing and Novel Strategies Against Algorithmic Discrimination Under Eu Law' Common Market Law Review 55 (2018) pp. 1143-1186.

[67] The DSA defines a "recommender system" as "*a fully or partially automated system used by an online platform to suggest in its online interface specific information to recipients of the service or prioritise that information, including as a result of a search initiated by the recipient of the service or otherwise determining the relative order or prominence of information displayed*" (art. 3/s)) and imposes, only on online platform providers, obligations of transparency of such systems (art. 27). In the case of providers of online platforms or very large online search engines, there are also duties to assess systemic risk, including assessing the "*design of their recommendation systems and any other relevant algorithmic system*" (art. 34/2/a)) and adopting measures to mitigate the risks identified in these systems (art. 35/1/d)). Under Article 38 of the DSA, very large online platforms and very large online search engines must allow users to configure recommendation systems so that they do not carry out profiling (a concept defined in Article 4/4 of the GDPR). Providers of these systems are also required to explain to regulators "*the design, logic, operation and testing of their algorithmic systems, including their recommendation systems*" (Art. 40/3 DSA). On the subject of recommender systems, see SERGIO GENOVESI / KATHARINA KAESLING / SCOTT ROBBINS (eds), *Recommender Systems: Legal and Ethical Issues* (Springer 2023) and MIREILLE HILDEBRANDT, 'The issue of proxies and choice architectures. Why EU law matters for recommender systems.' Frontiers in Artificial Intelligence 5 (2022): 789076.

**Transparency** can be understood as referring to several different concepts.[69] One of them, employed in Article 50, refers only to the identification of the origin of a given content or agent as being or coming from AI systems. Transparency is also covered by the obligation to provide and maintain technical documentation (arts. 11, 18, 20 and Annex IV), record-keeping (arts. 12 and 19), the provision of information (art. 13), and cooperation with authorities (art. 21).

When transparency refers to the characteristic of the AI system, this concept can allude to the description of the human tasks of designing, configuring and making the system available, even if the system is itself (i.e. in its operation) opaque. Transparency is sometimes used to refer to **interpretability**, i.e. the ability to understand how an AI system works,[70] and/or *explainability*, i.e. the clarification of why a certain result was obtained by operating the system.[71] A system can be interpretable, but produce concrete results that are not explainable.[72] For example, we know the parameters used and the steps followed by the system to assign a premium value in an insurance contract, but we can't explain why individual A has a higher premium than individual B. There are, however, artificial intelligence techniques that generate totally opaque systems (e.g.

---

[68] In particular in Art. 10 on data governance and Art. 15/4 on cybersecurity. The technical documentation required of providers of general purpose models also includes "*a detailed description of the elements of the model (…) and the relevant information on the development process, including (…) information on the data used for training, testing and validation, if applicable, including the type and provenance of the data and the curation methodologies (e.g. cleaning, filtering, etc.), the number of data points, their scope, etc, cleaning, filtering, etc.), the number of data points, their scope and main characteristics; how the data were obtained and selected, as well as all other measures to detect the inadequacy of data sources and methods to detect identifiable biases, if applicable*" (Annex XI, Section 1 (2)). On the other hand, the Regulation confers supervisory powers over high-risk AI systems on national public authorities or bodies that supervise or ensure compliance with obligations under Union law protecting fundamental rights, including the right to non-discrimination (Art. 77).

[69] The GDPR also uses the concept of transparency in art. 5/1 and recital 58, referring to the clear communication of information. Noting the "marked polysemy" of the concept of transparency, see Lorenzo Cotino Hueso, 'Transparencia y explicabilidad de la inteligencia artificial y "compañía" (comunicación, interpretabilidad, integilibilidad, auditabilidad, testabilidad, comprobabilidad, simulabilidad...). Para qué, para quién y cuánta.' in Lorenzo Cotino Hueso / Jorge Castellanos Claramunt (eds), *Transparencia y explicabilidad de la inteligencia artificial* (Tirant lo Blanch 2022) pp. 25 ff. In 2017 Zachary C. Lipton, *The Mythos of Model Interpretability*, arXiv:1606.03490 (2017) even stated: "*the term interpretability holds no agreed upon* meaning". The aforementioned 2020 technical standard used in this text seems to contribute to greater terminological certainty.

[70] This is the definition in the technical standard ISO/IEC TR 29119-11:2020(en), 3.1.42.

[71] See the definition used in the technical standard ISO/IEC TR 29119-11:2020(en), 3.1.31.

[72] Article 14/4/c) states that the system must allow a human being to "*correctly interpret the results of the high-risk AI system, taking into account, for example, the available interpretation tools and methods*". This wording seems to admit the use of so-called *black-box AI*, but in such cases there are no interpretation tools or methods available.

large language models, such as GPT); we know almost nothing about their inner workings.[73] For these, interpretability and explainability are not technically possible.[74]

The AI Act does not impose a general obligation to generate explainable models or decisions. However, in the case of high-risk systems, it establishes a right to an explanation of the role of the system (arts. 13 and 86), and to understand the main principles of its operation and the decision taken (arts. 14 and 86). The text of Article 86 (and recital 171) is not entirely clear as to whether it is necessary to explain the specific decision or whether a general explanation is sufficient.[75] On the other hand, the references to the relevant technical capacities to explain the results (Art. 13/3/b)/iv)) and "*where appropriate, information enabling those responsible for the deployment to interpret the results of the high-risk AI system and to use them appropriately*" (Art. 13/3/b)/vii)) are made in the context of technical documentation, which seems to indicate that a generic and abstract explanation (*interpretability*) is at stake and not a real *explainability*. Furthermore, even if a right to an explanation of the specific decision were established, the protection of personal data, business secrets and other types of secrecy would act as a limit to the exercise of this right.[76] In this sense, in my opinion, AI techniques that do not allow explanations to be generated (e.g. deep learning neural networks or *support vector machines*) remain legally admissible, even in the case of high-risk systems.

**Supervision and human control** are reflected in the obligation for the provider to adopt a risk management system (art. 9), quality control (art. 17), to monitor its post-marketing operation (art. 72), to report serious incidents (art. 73) and to design high-risk systems in a way that allows for understanding and intervention in their operation (art. 14), namely the existence of a kill switch (art. 14/4/e)). These aspects intersect with cybersecurity and robustness concerns (art. 15) - to which an important legislative framework is associated, namely the NIS 2 Directive (Dir. 2022/2555 of December 14, 2022 on measures for a high common level of cybersecurity across the Union) – and with the GDPR rule restricting the possibility of automated decisions to certain cases (art. 22 GDPR).[77]

---

[73] This is an area of scientific research. Recently, a large group of Anthropic researchers published a paper "Scaling Monosemanticity: Extracting Interpretable Features from Claude 3 Sonnet" (https://transformer-circuits.pub/2024/scaling-monosemanticity/index.html) in which the topic is discussed in detail and advances in the possibility of interpreting language models and using this technique for security purposes are demonstrated.

[74] Although recital 71 and, to a certain extent, art. 15/1/h) of the GDPR may give the impression that there would be a right to an explanation of automated decisions, this does not seem to be the most correct interpretation. See *supra* note 4s and also L. EDWARDS. / M. VEALE, 'Enslaving the algorithm: From a "right to an explanation" to a "right to better decisions"?' IEEE Security & Privacy, 16(3) (2018), pp.46-54.

[75] The different language versions (in English "meaningful explanation", in Portuguese "explicação clara e pertinente", in Spanish "claras y significativas", in French "claires et pertinentes", in Italian "chiare e significative" and in German "klare und aussagekräftige") are not conclusive.

[76] In a similar vein, see art. 25/5. There are also duties of secrecy and confidentiality (art. 78). On the wider problem see GIANCLAUDIO MALGIERI, 'Trade Secrets v Personal Data: A Possible Solution for Balancing Rights' International Data Privacy Law, vol. 6(2) (2016) pp. 102-116.

The implementation of these principles and of the Regulation will be densified to a large extent through *standards* and Commission guidelines, which will help to increase legal certainty.

**F. Prohibited practices**

At an early stage, the Commission proposed the establishment of four prohibited practices, said to pose an unacceptable risk, which could be summarily described as subliminal manipulation systems, systems that exploit vulnerabilities causing behavioral distortion and damage, social *scoring* systems and real-time biometric identification systems (e.g. facial recognition). These predictions had some exceptions and used particularly vague language.[78] After intense discussions and negotiations, the language has been refined, the list of prohibited practices has been extended, but the end result is not much better. They are now prohibited:

- Manipulation and exploitation of vulnerabilities – art. 5/1/a) and b)
- General *social scoring* – art. 5/1/c)
- Predictive policing – art. 5/1/d)
- Creation of facial recognition databases – art. 5/1/e)
- Emotion recognition systems in the workplace or education – art. 5/1/f)
- Biometric classification of protected categories – art. 5/1/g)
- Special cases of real-time biometric identification – art. 5/1/h)

This list is not exhaustive. Other practices may be prohibited or unlawful on other grounds (Art. 5/8). For example, systems that generate *deep fakes* are not normally seen as high-risk but are only subject to transparency obligations (Art. 50/4). However, when such a system is configured or prepared to generate child pornography that will be a crime[79]

---

[77] On this rule and the associated problems, see FEDERICO MARENGO, *Privacy and AI: Protecting Individual's Rights in the Age of AI* (2023).

[78] MICHAEL VEALE / FREDERIK ZUIDERVEEN BORGESIUS, (n 29), pp. 98-99: "*In briefings on the prohibitions, the Commission has presented an example for each. They border on the fantastical (...) A cynic might feel the Commission is more interested in prohibitions' rhetorical value than practical effect*".

[79] Curiously, the same might not necessarily be true of so-called "face swap porn" of adults. In Portugal, this practice has no clear criminal framework to date. For minors, Art. 176 of the Criminal Code is sufficient if there is a "realistic representation of a minor", regardless of whether there is a forgery involved. In the case of an adult, it is difficult to say that there is an offense against privacy (since there was no actual capture of real

## I. Manipulation and exploitation of vulnerabilities

A prerequisite for freedom in general, especially freedom of thought, choice and expression, is an adequate perception/representation of reality. Private autonomy and the free development of the personality require this. For this reason, the national legal system makes legal transactions concluded on the basis of defects of will voidable, and prohibits and punishes unfair commercial practices and misleading advertising. The autonomy of the will, as a reflection of the dignity of the human person, is also reflected in the prohibition of experimentation on people and the requirement of free and informed consent, especially in the case of voluntary limitation of personality rights.

Some AI systems have the potential to manipulate and mislead, interfering with the free formation of thoughts, opinions and choices.[80] In this sense, the text of art. 5/1/a) of the Regulation prohibits "*the placing on the market, the putting into service or the use of an AI system that deploys subliminal techniques beyond a person's consciousness or purposefully manipulative or deceptive techniques, with the objective, or the effect of materially distorting the behaviour of a person or a group of persons by appreciably impairing their ability to make an informed decision, thereby causing them to take a decision that they would not have otherwise taken in a manner that causes or is reasonably likely to cause that person, another person or group of persons significant harms.*" This wording uses indeterminate concepts and qualified language ("materially", "appreciably", "significant", "reasonably likely").[81] These qualifiers seem to indicate that not every advertising technique or hidden or misleading practice will be covered.[82] In fact, I believe that the criteria of advertising law and consumer protection will be less demanding, i.e., certain conduct qualified as aggressive or misleading advertising and/or commercial practices will not fall under Article 5/1/a) of the Regulation. In such cases,

---

images), and the problem can be seen from two perspectives: offense against image and good name. However, it doesn't seem to fall under the crimes of defamation, nor under the crimes of surveillance through the media, the Internet or other means of widespread public dissemination nor under illegal recordings and photographs. Article 5/1/b) of Directive 2024/1385 on combating violence against women and domestic violence seems to provide for the criminalization of this practice.

[80] Article 5/2 of the CoE Convention refers to the freedom to form opinions.

[81] There is controversy over the scientific basis of subliminal influence (i.e. that which falls below the threshold of conscious perception) of the subject. ROSTAM J. NEUWIRTH, 'Prohibited artificial intelligence practices in the proposed EU artificial intelligence act (AIA)' Computer Law & Security Review, 48 (2023), proposes the use of the term transliminal (instead of subliminal), since manipulation usually takes place between the plane of consciousness and unconsciousness.

[82] This overlaps with the topic of *dark patterns* (forms of user interface that promote an action or choice that users would be unlikely to make or take otherwise). On the subject *see* HARRY BRIGNULL, *Deceptive patterns – exposing the tricks tech companies use to control you* (Testimonium Ltd 2023) and INGE GRAEF, 'The EU Regulatory Patchwork for Dark Patterns: An Illustration of an Inframarginal Revolution in European Law?' (2023) https://ssrn.com/abstract=4411537.

the AI system will not be prohibited, but the activities in question, regardless of the use of an IT system, will be covered by the existing rules.

In turn, Article 5/1/b) prohibits "*the placing on the market, the putting into service or the use of an AI system that exploits any of the vulnerabilities of a natural person or a specific group of persons due to their age, disability or a specific social or economic situation, with the objective, or the effect, of materially distorting the behaviour of that person or a person belonging to that group in a manner that causes or is reasonably likely to cause that person or another person significant harm;* ". Such behavior, in the context of legal transactions, is already prohibited by contract and consumer law. Here, too, it seems that the qualifiers used and the limitation to certain characteristics make the Regulation's standard more demanding than the legislation in force, and, to that extent, the Regulation will have little impact. [83]

Let's think about personalized pricing systems that take into account that a potential customer is in a situation that makes them willing to pay a higher price (e.g. their cell phone is low on battery or their biometric data indicates dehydration or fatigue).[84] I believe that these situations would not fall within the scope of this article of the Regulation, although they could be considered illegal on other grounds.

**II. Social *scoring***

The practice of *scoring*, i.e. assigning numerical values to individuals, although not defined, is already covered by the GDPR, as it almost always involves profiling and often also an automated decision. This operation is often necessary so that computer systems can perform their functions. However, it raises concerns, especially considering what certain countries, such as India and China, have implemented: social classification systems, which take into account the generality of citizens' behavior in order to assign a classification that determines or influences their treatment in various contexts.[85]

---

[83] Rostam J. Neuwirth, (n 81), pp. 6-7. Vera Lúcia Raposo, 'Ex machina: preliminary critical assessment of the European Draft Act on artificial intelligence' International Journal of Law and Information Technology vol. 30 (2022) pp. 93-94.
[84] On the subject, mainly from an economic perspective, see Mateusz Grochowski / Fabrizio Esposito /Antonio Davola, *Price 'Personalization vs. Contract Terms Personalization: Mapping the Complexity* (2024) in https://ssrn.com/abstract=4791124.  irective (EU) 2019/2161 of the European Parliament and of the Council of 27 November 2019 amending Council Directive 93/13/EEC and Directives 98/6/EC, 2005/29/EC and 2011/83/EU of the European Parliament and of the Council in order to ensure better enforcement and modernization of Union rules on consumer protection, has imposed an obligation to provide information on whether prices are determined automatically.
[85] Cfr. Ralph Schroeder, 'Aadhaar and the Social Credit System: Personal Data Governance in India and China' International Journal of Communication vol. 16 (2022) pp. 2370-2386.

The Regulation only prohibits AI systems "*for the evaluation or classification of natural persons or groups of persons over a certain period of time based on their social behaviour or known, inferred or predicted personal or personality characteristics, with the social score leading to either or both of the following (...) detrimental or unfavourable treatment (...) in social contexts that are unrelated to the contexts in which the data was originally generated or collected (...)* [or] *that is unjustified or disproportionate to their social behaviour or its gravity*" (art. 5/1/c)). What is at stake is what is known as *general social scoring, i.*e. the overall assessment of a natural person's behavior.[86] On the other hand, AI systems that do more restricted *scoring,* such as those dedicated to *credit scoring*, solvency assessment or risk assessments and the pricing of life or health insurance, will be classified as high-risk (Annex III, 5/b) and c)).[87] Finally, systems that *score* for the purposes of detecting financial fraud or for setting prices in car insurance will not even be covered by the Regulation. Again, what determines the risk classification of the system is the purpose of the quantitative assessment and not the practice of *scoring*.

As has been pointed out, *scoring* is usually associated with an automated decision, which, when involves the processing of personal data and produces effects on the legal sphere or significantly affects the personal data subject, may from the outset be prohibited under Article 22 GDPR.[88] However, it is important to note that Article 22 of the GDPR only applies to *fully* automated decisions.[89] Therefore, at least in the case of high-risk systems, where the regulation requires human supervision (Art. 14 of the Regulation), it is possible to escape the application of this GDPR rule.

---

[86] Nizan Geslevich Packin, 'Disability Discrimination Using Artificial Intelligence Systems and Social Scoring: Can We Disable Digital Bias?' Journal of International Comparative Law (2021) p. 496: "*Social scoring, however, attempts to systematically rate people in their entirety (and not just their creditworthiness) based on social, reputational and even behavioral features (as opposed to credit history)*". On the phenomenon see Danielle Keats Citron / Frank Pasquale, 'The Scored Society: Due Process for Automated Predictions' Washington Law Review 89 (2014) pp. 1-33.
[87] Vera Lúcia Raposo, (n 83) p. 94 points out that the reference to "a certain period of time" will exclude episodic scoring.
[88] In judgment C-634/21, *Schufa*, (EU:C:2023:957), §44-46 the ECJ adopted a broad concept of decision, saying that a *credit score* qualified as such.
[89] The standard requires " *...three cumulative conditions, namely, first, that there must be a 'decision', secondly, that that decision must be 'based solely on automated processing, including profiling', and, thirdly, that it must produce 'legal effects concerning [the interested party]' or 'similarly significantly [affect] him or her'..*" (C-634/21, *Schufa,* §43). The EDPB, Guidelines on Automated individual decision-making and Profiling for the purposes of Regulation 2016/679 (2018) p. 21 point out that merely symbolic human intervention is not enough. A decision is not considered fully automated when there are organizational measures that ensure substantial and structured human involvement. In case law, see the decision of the Rechtbank Amsterdam of 11.III.2021 (ECLI:NL:RBAMS:2021:1018,) in which the requirement of a consensus between several people was at issue), the decision of the Rechtbank Den Haag of 11.II.2021 (NL:RBDHA:2020:1013) in which a right of veto was provided for and a decision of the Austrian Bundesverwaltungsgericht of 18.XII.2020 (AT:BVWG:2020:W256.2235360.1.00) in which there were training and guidelines for dealing with the recommendation produced by the system. For more case law see Sebastião Barros Vale / Gabriela Zanfir-Fortuna, *Automated Decision-Making Under the GDPR: Practical Cases from Courts and Data Protection Authorities* (Future of Privacy Forum 2022).

**III. Biometric identification and classification, including sentiment detection**

Biometric identification systems, especially those for facial and emotion recognition, generated significant controversy during the legislative process. From the outset, these systems constitute an attack on individual privacy and freedom, with a high discriminatory potential. In this sense, companies such as Clearview.AI, which systematically scraped the Internet (especially social networks) to generate a facial recognition database, had already been sanctioned for violating the GDPR.[90] In any case, the Regulation now expressly prohibits this practice (art. 5/1/e)).

The use of emotion recognition systems has been challenged on technical grounds. It is argued that expressions are variable at an individual level and depend on the social and cultural context, so these systems are not reliable. In addition, they have a high discriminatory potential.[91] Paradoxically, the Regulation only prohibits the use of these emotion recognition systems in the context of work and education. In all other cases, emotion recognition systems are considered high-risk systems (Annex III/1/c). Both teaching and work can be done remotely, but I believe these situations are covered by the ban. On the other hand, the ban does not cover "*AI systems placed on the market strictly for medical or safety reasons, such as systems intended for therapeutical use*". This will raise questions in cases where systems are used for safety or medical reasons in the areas of workplace and education institutions. In that scenario, the intention seems to be allowing the use of such systems. Automatic interview systems should be classified as high-risk (Annex III,4), unless they also include an emotion recognition component.[92]

On the other hand, the very notion of emotion recognition must be read restrictively. Recital 18 explains: "*The notion refers to emotions or intentions such as happiness, sadness, anger, surprise, disgust, embarrassment, excitement, shame, contempt, satisfaction and amusement. It does not include physical states, such as pain*

---

[90] The company was subject to fines of 20 million euros in France (2021, there was also a penalty payment of five million in 2023), Greece (2022) and Italy (2022). In 2023, the Austrian authority also considered this company's activity to be in breach of the GDPR, but did not impose any fines or other measures. In 2021, the Swedish supervisory authority fined police authorities for using Clearview's services. On the other hand, in the UK, the same company succeeded, in a court decision of 17.X.2023, in overturning the fine imposed, based on a question of jurisdiction and applicable law, particularly in light of *Brexit* – [2023] UKFTT 00819 (GRC).
[91] See recital 44.
[92] With a very critical view of these systems see IFEOMA AJUNWA, 'Automated video interviewing as the new phrenology' Berkeley Technology Law Journal vol. 36 (2021) pp.1173-1225.

*or fatigue, including, for example, systems used in detecting the state of fatigue of professional pilots or drivers for the purpose of preventing accidents. This does also not include the mere detection of readily apparent expressions, gestures or movements, unless they are used for identifying or inferring emotions. Those expressions can be basic facial expressions, such as a frown or a smile, or gestures such as the movement of hands, arms or head, or characteristics of a person's voice, such as a raised voice or whispering.*".

While biometric identification in public spaces can serve laudable purposes (e.g. finding missing persons or fugitives), its operation implies the compression of citizens' privacy and the creation of a state of constant surveillance, intolerable in a democracy with European values.[93] In this sense, in 2023, in a unanimous decision, the ECtHR confirmed that the use of facial recognition technology to identify, locate, and arrest an individual in an administrative offense proceeding was unlawful (in violation of Article 8 of the ECHR).[94]

The solution adopted in Article 5/1/h) of the Regulation is to prohibit the use of these systems of "*real–time' remote biometric identification systems in publicly accessible spaces for the purposes of law enforcement* ", except when strictly necessary for one of three objectives: "(*i) the targeted search for specific victims of abduction, trafficking in human beings or sexual exploitation of human beings, as well as the search for missing persons; (ii) the prevention of a specific, substantial and imminent threat to the life or physical safety of natural persons or a genuine and present or genuine and foreseeable threat of a terrorist attack; or (iii)the localisation or identification of a person suspected of having committed a criminal offence, for the purpose of conducting a criminal investigation or prosecution or executing a criminal penalty for offences referred to in Annex II and punishable in the Member State concerned by a custodial sentence or a detention order for a maximum period of at least four years.*" In such cases, Art. 5/2 requires a fundamental rights impact assessment (Art. 27) and registration (Art. 49), and Art. 5/4 specifies that the relevant market surveillance and data protection authorities must be4 notified of such use.

---

[93] This matter is already regulated by Directive 2016/680. On the subject cfr. Vera Lúcia Raposo 'Look at the camera and say cheese': the existing European legal framework for facial recognition technology in criminal investigations' Information & Communications Technology Law, 33(1) (2024) pp. 1-20.
[94] *Glukhin v. Russia*, 11519/20 (decision of 4.VII.2023).

It should be noted that remote biometric identification for other purposes or on a delayed basis is not prohibited,[95] and is generally classified as a high-risk use, except in the case of simple identity recognition and verification systems (Annex III, 1, a)).[96]

The Regulation also deals with biometric *categorization*, which differs from biometric *identification*. While in identification the aim is to determine who the person is, starting from certain physical, psychological or behavioral characteristics (biometric data - art. 3/34) to arrive at an individual; biometric categorization aims to classify the subject - to know if someone has a given characteristic.[97] Thus, in biometric *identification*, the system will know from my face that I am Nuno Silva, in biometric *categorization, from the* way I walk, the system will determine whether I have a risk of developing Alzheimer's or, by analyzing my face, it will assess whether I am a dangerous anarcho-syndicalist.

According to the Regulation, biometric categorization systems "*that categorise individually natural persons based on their biometric data to deduce or infer their race, political opinions, trade union membership, religious or philosophical beliefs, sex life or sexual orientatio*n" are prohibited (art. 5/1/g)).[98] Thus, systems such as the controversial neural network that allegedly detected people's sexual orientation from photographs will not be admissible.[99] There is, however, a caveat for processing and categorizing biometric data in the field of law enforcement, which remains admissible.[100] On the other hand, "*AI systems intended to be used for biometric categorisation, according to sensitive or protected attributes or characteristics based on the inference of those*

---

[95] Article 26/10 states that, in the case of post-remote biometric identification systems (defined in Article 3/43, as opposed to "real-time" systems defined in Article 3/42), "*the deployer* (...) shall *request an authorisation, ex ante, or without undue delay and no later than 48 hours, by a judicial authority or an administrative authority whose decision is binding and subject to judicial review, for the use of that system, except when it is used for the initial identification of a potential suspect based on objective and verifiable facts directly linked to the offence. Each use shall be limited to what is strictly necessary for the investigation of a specific criminal offence*" If authorization is rejected, use must cease and the data must be destroyed. It also prohibits indiscriminate use ("non-selective") and allows member states to adopt more restrictive legislation.

[96] See recitals 15, 17 and 52 and the definition of biometric verification (art. 3/36).

[97] Biometric categorization system is defined in Art. 3/40 as "*an AI system for the purpose of assigning natural persons to specific categories on the basis of their biometric data, unless it is ancillary to another commercial service and strictly necessary for objective technical reasons.*" (for examples of ancillary categorization see Recital 16), while biometric identification concerns the "*automated recognition of physical, physiological, behavioural, or psychological human features for the purpose of establishing the identity of a natural person by comparing biometric data of that individual to biometric data of individuals stored in a database*" (Art. 3/35).

[98] This provision can be criticized for being too restricted in the "protected categories".

[99] The controversial original study has been replicated by JOHN LEUNER, 'A replication study: Machine learning models are capable of predicting sexual orientation from facial images' *arXiv:1902.10739* (2019), who argues that these models take into account other factors and not facial physiognomy/structure.

[100] See Recital 30.

*attributes or characteristics*" are not prohibited, but are classified as high-risk systems (Annex III, 1, b)).

**IV. Predictive policing**

The definition of profiles is based on the repeatability and standardization of behaviour. It is based on the idea that the past repeats itself in the future and that there are certain features of individuals that have predictive capacity. The application of these techniques in the criminal context raises special concerns, especially given the potential consequences of an error or injustice and the presumption of innocence.[101]

Thus, the Regulation prohibits predictive policing practices that use AI systems to assess the risk of a natural person committing a criminal offense "*based solely on the profiling of a natural person or on assessing their personality traits and characteristics*" (art. 5/1/d)).

However, "*this prohibition shall not apply to AI systems used to support the human assessment of the involvement of a person in a criminal activity, which is already based on objective and verifiable facts directly linked to a criminal activity*." In other words, the system must consider the concrete behavior and particular traits of a specific person and not their membership in certain categories or groups. This exception recognizes the potential usefulness of AI in the context of criminal investigation and prevention while ensuring that the assessment is based on actual data and not *exclusively* on (necessarily speculative) profiling.

In fact, predictive policing can be geared towards predicting crimes, predicting or identifying criminals, and/or predicting or identifying potential victims of crime.[102] Most of these systems, when not based exclusively on profiling, will fall under the high-risk classification (Annex III, 6). In this vein, recital 42 clarifies that the prohibition of Article 5/1/d) does not cover "*AI systems using risk analytics to assess the likelihood of financial fraud by undertakings on the basis of suspicious transactions or risk analytic tools to predict the likelihood of the localisation of narcotics or illicit goods by customs authorities, for example on the basis of known trafficking*".

---

[101] As can be read in recital 42: " *In line with the presumption of innocence, natural persons in the Union should always be judged on their actual behaviour. Natural persons should never be judged on AI-predicted behaviour based solely on their profiling, personality traits or characteristics, such as nationality, place of birth, place of residence, number of children, level of debt or type of car, without a reasonable suspicion of that person being involved in a criminal activity based on objective verifiable facts and without human assessment thereof.*"
[102] WALTER PERRY et al, *Predictive Policing: The Role of Crime Forecasting in Law Enforcement Operations* (RAND Corporation 2013).

A well-known example of an AI system for predictive purposes in the criminal context is the COMPAS (*Correctional Offender Management Profiling for Alternative Sanctions*) system, used in some US courts to calculate the risk of recidivism and, on that basis, define sentencing.[103] Tools like this, if they are not based exclusively on profiling, are not covered by the ban but are considered high-risk AI systems (Annex III, 6 d) and e) and 8)).

## G. High-risk systems
### I. Qualification

The definition of high-risk systems is made in article 6 by reference to two Annexes.[104]

Annex I includes legislation on certain categories of products (such as toys, vehicles, explosives, elevators, or medical devices) and, according to Article 6/1, when AI systems are used as safety components in these products (or the AI systems are themselves products[105]) subject to a conformity assessment obligation, this is a high-risk system.[106]

In turn, Article 6/2 refers to Annex III, which specifies certain uses such as biometric identification, management of critical infrastructures, admission and classification in educational establishments, job interviews, monitoring of workers, access to and use of essential services (public and private), use in border control, in a judicial context or by law enforcement agencies. As PHILIP HACKER points out,[107] more important than the context of use is the purpose - a system used for medical operations or triage does not carry the same risk as a system that manages medical appointments.

The system works with auto-classification, i.e. each operator will determine the risk classification of their system. It is important to read the various hypotheses carefully and consider the Regulation's recitals. The Commission will adopt guidelines specifying the practical application of this article "*together with a comprehensive list of practical examples of cases of use of high-risk and non-high-risk AI systems*" (art. 6/5).

---

[103] The subject of much academic and judicial discussion. In the well-known *Loomis v. Wisconsin,* 881 N.W.2d 749 (Wis. 2016), cert. denied, 137 S. Ct. 2290 (2017) the Wisconsin Supreme Court rejected the appeal of an individual who had been considered by the software to have a high-risk of recidivism and thus sentenced to 6 years in prison. According to the court, *due process* had not been violated despite the fact that the sentence had been determined using COMPAS, whose algorithm and mode of operation is unknown. The discriminatory nature of this system was the subject of a report by ProPublica.
[104] The reason for this definition being made by reference is to make it easier to update these annexes in the simplified procedure (delegated acts of the European Commission) provided for in Articles 6/6, /7 and /8 and 7.
[105] This can happen in particular with toys or medical devices.
[106] "Safety component" is defined in Article 3/14 as "*a component of a product or of an AI system which fulfils a safety function for that product or AI system, or the failure or malfunctioning of which endangers the health and safety of persons or property."* This definition is broad, but it should be read using a normality/predictability criterion.
[107] *AI Regulation in Europe: From the AI Act to Future Regulatory Challenges* (2023) arXiv:2310.04072 p. 7.

The risk classification is based on the intended use, but there are some caveats. For example, remote biometric identification systems are generally high-risk, but there is an exclusion for identity verification systems (Annex III(1)(a)). Similarly, systems for assessing the *creditworthiness of* natural persons or *credit scoring* are high-risk systems, except when such systems are used for the detection of financial fraud (Annex III(5)(b)).

In addition to specific exceptions, there is a more general derogation. According to Article 6/3, it is possible to disregard the high-risk classification for a system whose foreseeable use is listed in Annex III "*if it does not pose a significant risk of harm to the health, safety or fundamental rights of natural persons, including by not materially influencing the outcome of decision making*" and provided that it does not carry out profiling of natural persons (last paragraph of this Article 6/3). However, this does not mean that all profiling AI systems are deemed high-risk. All that it means is that *prima facie* high-risk systems that carry out profiling will not be able to invoke the exemption.

Article 6/3 sets out circumstances in which such AI systems, despite having a purpose set out in Annex III, will not pose a significant risk: a) when they perform a narrow procedural task; b) when they are intended to improve the result of a previously completed human activity; c) when they aim to detect decision-making patterns or deviations from previous decision-making patterns and are not intended to replace or influence a previously completed human assessment; or d) when the AI system is intended to perform (only) a preparatory task. For an AI system not to be considered high-risk, despite its intended purpose, it is sufficient to meet one of these points and not to carry out profiling (as defined in Art. 4/4 GDPR).

Recital 53 gives some examples of such systems, including AI systems designed to improve the language used in previously drafted documents, for example in relation to professional tone, academic style or the alignment of the text with a particular brand message, systems that are used to check *ex post* whether a teacher may have deviated from their usual pattern of awarding marks, in order to flag up potential inconsistencies or anomalies, intelligent file handling solutions that include various functions such as indexing, searching, text and voice processing or linking data to other data sources, or AI systems used for document translation.

In any case, anyone wishing to invoke this derogation must document this assessment (Art. 6/4) and register it (Art. 49/2). A market surveillance authority may, however, disagree and demand corrective action (Art. 80).

**II. Rules**

In simple terms, the Regulation requires high-risk systems to be well-made, properly maintained, and controlled. The operators must have adequate documentation to prove compliance with the Regulation's rules.

Machine learning systems are subject to data quality requirements, particularly in terms of representativeness and the application of measures to detect and mitigate *biases* (Art. 10). Article 10(5) even creates a new basis for the lawful processing of sensitive data (in addition to those in Article 9 of the GDPR) by establishing that, under specific conditions, it will be possible to process special categories of personal data "*to ensure bias detection and correction* ".[108] On the other hand, most of the Regulation's provisions will legitimize the processing of non-sensitive data since this will occur in order to comply with legal obligations (art. 6/1/c) GDPR).[109]

Providers of high-risk systems are responsible for meeting the requirements of articles 8 to 15 (art. 16), as well as ensuring the existence of a quality management system (art. 17), keeping documentation for a period of 10 years after the system has been placed on the market or put into service (art. 18), and maintaining logs (art. 19). There is also a duty to cooperate with competent authorities (articles 20/2, 21 and 73), to adopt corrective measures (article 20/1), and perform post-market monitoring (article 72). This monitoring includes a duty to inform the authorities in the event of a serious incident (Art. 73), defined in Art. 3/49 as "*any incident or malfunctioning in an AI-system which, directly or indirectly, has any of the following consequences: (a) death of a person or serious harm to a person's health (b) a serious and irreversible disruption of the management or operation of a critical infrastructure, (c) infringement of obligations under Union law designed to protect fundamental rights, (d) serious harm to property or the environment*".

From a more bureaucratic point of view, in addition to a duty of documentation and record-keeping, providers of high-risk AI systems are obliged to identify themselves as such (art. 16/b)) and to follow a conformity assessment procedure (art. 43),[110] including drawing up a declaration of conformity (art. 47), using the CE mark (art. 48) and registering the high-risk system (arts. 49 and 71).[111]

---

[108] This may make it difficult to apply bias mitigation measures to systems that are not high-risk, since for these there will be no lawful basis for processing sensitive data (MICHAEL VEALE / FREDERIK ZUIDERVEEN BORGESIUS, (n 29), p. 103). Additional processing of personal data is also provided for under certain conditions to safeguard the public interest (Art. 59).

[109] There is no equivalent basis for sensitive data, hence the need for article 10(5) of the AI Act.

Although the most important duties fall on the *providers of* AI systems, their *users* ("deployers") are also subject to several obligations set out in Article 26. To the extent that they control the system, deployers will have to respect the instructions for use of the system, ensure its human supervision and the quality and appropriateness of the input data, collaborate with the authorities, keep records of the system's operation and inform natural persons that they are subject to the use of the high-risk AI system.

In some cases, bodies governed by public law or private entities providing public services, as well as banks and insurance companies, must carry out a fundamental rights impact assessment (Art. 27). This assessment is not to be confused with the obligation to carry out a data protection impact assessment (*DPIA*) laid down in Article 35 of the GDPR, although the Regulation itself recognizes the existence of partial overlaps (Article 27/6).

**H. Obligation of transparency for certain systems**

Article 50 of the Regulation, the only one in Chapter IV, deals with certain systems defined in the light of their purpose, imposing minimum transparency/information requirements.[112] The first two paragraphs of this article impose duties on providers, while paragraphs 3 and 4 concern the duties of those responsible for implementing these AI systems.[113] These duties apply to the systems mentioned in Article 50 regardless of their risk classification.

Article 50/1 regulates AI systems "*intended to interact directly with natural persons*", i.e. so-called chatbots or conversational systems. These systems must be designed in such a way that it is clear to natural persons "*that they are interacting with an AI system, unless this would be obvious from the point of view of a natural person*

---

[110] A derogation from this procedure is provided for, particularly in cases of urgency (Article 46).
[111] Taking into account the principles of country of origin and mutual recognition, this operation only needs to be carried out in one Member State.
[112] The duties of transparency/disclosure set out in Article 50 do not apply when the system is legally authorized "*to detect, prevent, investigate or prosecute criminal offences, subject to appropriate safeguards for the rights and freedoms of third parties* ".
[113] It is unclear whether the manufacturers of these systems are covered by the exemption from liability in Article 6 of the DSA. From the outset, it is debatable whether we can classify providers of general-purpose or generative AI models or systems as an "intermediary service" (as provided for in art. 3/g) of the DSA). Recital 119 of the AI Act seems to point to a case-by-case assessment.

*who is reasonably well-informed, observant and circumspect, taking into account the circumstances and the context of use* ".

Generative AI systems ("*generating synthetic audio, image, video or text content*") are addressed in Article 50/2. There is an obligation to identify such synthetic content with a digital "watermark" "*in a machine-readable format and detectable as artificially generated or manipulated*".[114]

Those responsible for implementing emotion recognition or biometric categorization systems are subject to a duty to disclose its use (art. 50/3).[115] Similarly, those who create *deepfakes* must "disclose *that the content has been artificially generated or manipulated*" (Art. 50/4 1st paragraph). [116] This duty can be compressed "*where the content forms part of an evidently artistic, creative, satirical, fictional or analogous work or program*". In that scenario; it is sufficient that the disclosure is done in " *an appropriate manner that does not hamper the display or enjoyment of the work*.". The duty of disclosure also exists in the case of news ("*text which is published with the purpose of informing the public on matters of public interest* "), except " *AI-generated content has undergone a process of human review or editorial control and where a natural or legal person holds editorial responsibility for the publication of the content* " (art. 50/4/2nd paragraph).

**I. General purpose models**

When the European Commission presented the proposal for a Regulation in April 2021, there were already some AI models with diversified capabilities, but the term "foundational models", used to indicate those models trained with large amounts of data and with the potential for various applications, had not yet been coined. It wasn't until August 2021 that a paper by Stanford researchers used this notion for the first time.[117] The real explosion of foundational models, which include GPTs from the OpenAI company and competitors PALM, BERT and Gemini (Google), Claude (Anthropic), Luminous (Aleph Alpha), Mistral 7B and LlaMA (Meta) took place in 2023.

---

[114] Watermarking implementation solutions must provide technical solutions that are " *effective, interoperable, robust and reliable as far as this is technically feasible, taking into account the specificities and limitations of various types of content, the costs of implementation and the generally acknowledged state of the art, as may be reflected in relevant technical standards* ". This obligation does not apply to editing support tools (such as a spelling checker) and in general those that "*do not substantially alter the input data provided by the deployer or the semantics thereof*" (Art. 50/2).

[115] As we have seen, this type of system can be banned or classified as high-risk. In any case, as MICHAEL VEALE / FREDERIK ZUIDERVEEN BORGESIUS point out, (n 29) p. 107, this duty does not seem to add anything to what already results from the GDPR.

[116] To use the expression of MICHAEL VEALE / FREDERIK ZUIDERVEEN BORGESIUS, (n 29). However, as they point out, a teleological understanding of this obligation should except uses in contexts where there is no risk of deception (as in the case of generic images used for marketing or presentation purposes). Recitals 132 and 133 seem to support this interpretation.

[117] RISHI BOMMASANI et al, *On the Opportunities and Risks of Foundation Models*, arXiv:2108.07258 [cs.LG].

This technology has particularities that are especially challenging. On the one hand, they have high development costs, which create considerable barriers to entry. Unlike the specialized systems for which the Regulation was initially intended, these models have a capacity for generalization and will often be made available through programming interfaces (APIs) so that third parties can optimize and adapt them to specific applications. In this sense, these models, as ANDREJ KARPATHY explains,[118] are close to operating systems, generating considerable dependencies. These considerations are typically addressed by Competition Law,[119] but the Regulation has dedicated a chapter to them. Articles 89/2 and 93 provide for the protection of downstream providers, i.e., those who integrate a general-purpose model or system into their system and who become dependent on a general-purpose system that they do not control.

On the other hand, these large general-purpose models are often opaque: they are a vast array of numbers (the so-called parameters and weights of a neural network) that interact in ways that are beyond human comprehension. This lack of understanding raises concerns of security, control, and alignment.

In addition, developing these models requires massive amounts of data, much of which is taken from the Internet and includes personal data and data protected by intellectual property rights. Furthermore, contrary to what was initially thought, these models retain some of the data in "memory".[120] This makes assessing the lawfulness of these uses even more complex.

Finally, most foundational models have "creative" capacities and also fall into the category of generative AI covered by Article 50.[121]

The Regulation deals with all general purpose AI models (in Articles 53 and 54) and imposes additional duties (in Article 55) for so-called general purpose AI models with systemic risk.[122] According to Article 51, systemic risk exists if the model has "*high*

---

[118] This statement is made in several public lectures available on Youtube. I especially suggest the video "[1hr Talk] Intro to Large Language Models".
[119] Cf. HOU LIYANG, 'The Essential Facilities Doctrine – What was Wrong in Microsoft?' IIC 43(4) [2012] pp. 251-271.
[120] MILAD NASR et al, *Scalable Extraction of Training Data from (Production) Language Models*, arXiv:2311.17035 [cs.LG].
[121] Not all generative AI systems are foundational models; there are a number of specialized applications for creating music, images, text, etc.
[122] As already mentioned, general purpose AI models are defined in Art. 3/63. "Systemic risk", in turn, is defined as "*a risk specific to the high-impact capabilities of general purpose AI models that have a significant impact on*

*impact capabilities evaluated on the basis of appropriate technical tools and methodologies, including indicators and benchmarks* " (51/1/a)) or equivalent capabilities or impact taking into account the criteria set out in Annex XIII, on the basis of a decision by the Commission, ex officio or following a qualified alert by the scientific pane*l*" (51/1/b)). "High impact capabilities" is defined as "*capabilities that match or exceed the capabilities recorded in the most advanced general purpose AI models*" (Art. 3/64). In other words, in this matter, the legislator essentially refers to technical-scientific criteria set out in Annex XIII, and which will be fleshed out by the Commission in delegated acts (Art. 51/3). In any case, Article 51(2) establishes a (rebuttable) presumption that the model has high impact capabilities when the cumulative amount of computation used for its training, measured in floating point operations per second (FLOPS), is greater than $10^{25}$ . [123]

Article 52 sets out the procedure for classifying a model as having systemic risk, in which the provider "*may present (…) sufficiently substantiated arguments to demonstrate that, exceptionally, although it meets this requirement, the general purpose AI model does not present, due to its specific characteristics, systemic risks and, therefore, should not be classified as a general purpose AI model with systemic risk*" (Article 52/2).

Providers of general-purpose AI models are essentially subject to four duties set out in Article 53: i) to maintain appropriate and up-to-date technical documentation (paragraph 1/b) and Annex XI); ii) to facilitate integration and interoperability with their system (paragraph 1/b and, Annex XII); iii) to apply a policy of respect for copyright (paragraph 1/c)), in particular ensuring that the system respects the reservation of rights provided for in Article 4 of Directive 2019/790 in the context of *text and data mining*[124] and (iv) make publicly available a summary of the content used to train the model (paragraph 1/d), which provides for a model of this summary to be drawn up by the AI Office). In addition, there is a general duty to cooperate with the authorities (Art. 53/3).

---

*the Union market due to their reach or due to actual or reasonably foreseeable negative effects on public health, safety, public security, fundamental rights or the society as a whole, that can be propagated at scale across the value chain*" (Art. 3/65).
[123] Floating-point operations are defined in Article 3/67 as "*any mathematical operation or assignment involving floating-point numbers, which are a subset of the real numbers normally represented in computers by an integer of fixed precision scaled by an integer exponent of a fixed base*". In this context, this value is a measure of the performance and computational capacity of the *hardware* used to train a given AI model. The higher it is, the greater the complexity of the models and the corresponding training costs. Interestingly, the US Executive Order uses 10^26 FLOPS as the threshold, i.e. ten times more.
[124] The Regulation devotes recitals 105 to 108 to the subject of copyright. See ALEXANDER PEUKERT, 'Copyright in the Artificial Intelligence Act - A Primer' GRUR-Int vol. 73(6) (2024) pp. 497-509.

In the case of models with systemic risk, in addition to the duties applicable to all general purpose models, Article 55(1) stipulates that the respective providers must: a) carry out tests and evaluations of the model with a view to identifying and mitigating systemic risks; b) assess and mitigate any of those risks; c) monitor, document and communicate relevant information on serious incidents and any corrective measures to resolve them; and d) ensure an adequate level of protection in terms of cybersecurity.

If the model providers are established in third countries (outside the EU), an authorized representative will primarily carry out these duties, as established in Article 54.

**J. Certification, supervision, and sanctions**

The Regulation establishes preventive and repressive measures, although it essentially focuses on the placing on the market or putting into service of high-risk AI systems. Although civil liability is not directly addressed,[125] some of the Regulation's rules if breached, could give rise to an indemnifying obligation under national rules. In addition, there is a reference to the possibility of collective claims under the terms of Directive 2020/1828 (art. 110).

Since this is product safety legislation, articles 28 ff. provide for a certification and control scheme. There will be at least one national notifying authority[126] and a national market surveillance authority (art. 70), which will be the competent national authorities under the terms of the Regulation.[127]

The notifying authority is the one that assesses, designates, and supervises the conformity assessment bodies: typically, independent private entities that carry out testing, certification, and inspection activities on the systems to ensure that they meet the requirements of the Regulation. Notified bodies are a special category of officially designated conformity assessment bodies with CE marking competence.[128]

---

[125] As mentioned, this issue is addressed in two Directives still at the proposal stage: COM(2022)495 final and COM(2022)496 final.
[126] The definition of notifying authority ("*the national authority responsible for setting up and carrying out the necessary procedures for the assessment, designation and notification of conformity assessment bodies and for their monitoring*") is set out in article 3/19.
[127] Arts. 3/48 and 74. Different Member States have taken different approaches. Some, like Spain, have created a new authority. Others have preferred a decentralized system, using only sectoral regulators. Some have sought to assign these powers to existing authorities, such as the supervisory authorities in the field of data protection or the digital services coordinators under the DSA. In the case of EU activities subject to the Regulation, the supervisory authority will be the European Data Protection Supervisor (Art. 74/9), who will also have the power to impose fines (Art. 100).

As provided for in Articles 40 and 42, the European standardization organizations will develop standards that will be adopted by the European Commission under Regulation (EU) No 1025/2012. Following these standards in a high-risk AI system will give rise to a presumption of conformity (arts. 40/1 and 42).[129]

National market surveillance authorities will deal with complaints (Art. 85) and serious incidents (Art. 73) and exercise the powers provided for in Regulation 2019/1020 (Art. 74), including risk assessments, imposing corrective measures (Art. 79), detecting non-compliance (Art. 83) and supervising tests in real conditions (Art. 76). It is also expected that these will be the authorities with sanctioning powers.

At European level, the Commission, through its *AI Office* (arts. 3/47 and 64),[130] will supervise general-purpose AI models, functioning for this purpose as a market surveillance authority (art. 75), with extensive supervisory powers (arts. 88 to 94) and the power to impose fines (art. 101). In addition to the AI Office, there is also a European AI Board (Art. 65), made up of a representative from each Member State, whose main function is to coordinate the application of the Regulation between the various States (Art. 66). The AI Service and the Committee will be assisted by an advisory forum (Art. 67) and a scientific panel of independent experts (Art. 68).[131]

Sanctions vary according to the type of infringement, must take into account the specific circumstances (Art. 99/7), and may include warnings and non-pecuniary measures (Art. 99/1).[132] There are fines of up to 7% of worldwide turnover or 35 million euros in the case of prohibited practices (art. 99/3), up to 3% of turnover or 15 million euros for general infringements (art. 99/4) and up to 1% or 7.5 million euros in the case of providing "incorrect, incomplete or misleading" information to notified bodies and competent authorities (art. 99/5).[133]

---

[128] See art. 3/21 and /22 and in more detail the *2022 Blue Guide on the application of EU rules on products* (2022/C 247/01). The European Commission maintains a list of notified bodies, known as NANDO (https://webgate.ec.europa.eu/single-market-compliance-space/#/notified-bodies).

[129] The European standardization bodies are the European Committee for Standardization (CN), the European Committee for Electrotechnical Standardization (CENELEC) and the European Telecommunications Standards Institute (ETSI). There is also provision for the Commission to adopt common specifications if these organizations fail (Article 41). The request to issue standards relating to this Regulation was already submitted by the Commission to the CN and CENELEC in May 2023 (C(2023)3215 – Standardization request M/593). On the process and the role of standards in the Regulation *see* MARTA CANTERO GAMITO / CHRISTOPHER T MARSDEN, 'Artificial intelligence co-regulation? The role of standards in the EU AI Act' International Journal of Law and Information Technology, vol. 32 (1) (2024).

[130] This department of the European Commission was created by Commission Decision of January 24, 2024 (C(2024) 390 final).

[131] On this institutional framework see CLAUDIO NOVELLI et al, 'A Robust Governance for the AI Act: AI Office, AI Board, Scientific Panel, and National Authorities' (2024) at https://ssrn.com/abstract=4817755.

[132] Although the Regulation does not expressly mention it, it seems that the broad understanding of "company" from Competition Law, which has been used in regulation, namely in data protection law and digital platforms, especially for sanctioning purposes, should apply. The definition is "*any entity engaged in an economic activity, regardless of the legal status of that entity and its method of financing*" (see C-138/11, *Compass-Datenbank GmbH*, EU:C:2012:449, §35).

The fact that an entity is sanctioned under the Regulation does not prevent other fines from being imposed, namely for violating the GDPR or the DSA.

**K. Conclusion**

In my opinion, the Regulation contains generally balanced and reasonable solutions. However, given its length, complexity, and poor legislative quality, it will become somewhat difficult to implement.[134] There is, therefore, a real risk that the European Union will negatively affect innovation and investment in the field of Artificial Intelligence. It is also possible that there will be a reduction in the supply and/or divergence of products or services, with the European public receiving different and less advanced versions.[135] As MIGEL PEGUERA POCH writes,[136] the Regulation is a remarkably complex instrument with unpredictable effects.

The main hope lies in the use of standards, whose mass adoption could significantly reduce *compliance* costs and reduce the considerable uncertainty that this legislative instrument will inevitably generate.[137] Another contribution to overcoming the limitations of this piece of legislation will have to come from lawyers.

---

[133] For providers of general purpose AI models, the framework is the same (Art. 101). Interestingly, in the case of European authorities, the maximum amount is only 1.5 million euros for prohibited practices (Art. 100/2) and 750,000 euros in other cases (Art. 100/3). More important is the possibility given to Member States to "define rules to determine the extent to which fines may be imposed on public authorities and bodies established in that Member State." (art. 99/8). In other words, as with the GDPR, it seems legally permissible to exempt public bodies from fines. The best example comes from above…

[134] MICHAEL VEALE / FREDERIK ZUIDERVEEN BORGESIUS, (n 29)

[135] LUCIANO FLORIDI, (n 43) : "*fridges, dishwashers, washing machines and even vehicles may need to remain on the safe side of "artificial stupidity" to avoid having to comply with the AI Act (CP version). A scenario becomes plausible in which companies start dumbing down ("de-AI-ing") or at least stop smartening up their products in order not to be subject to the AI Act*.". This does not appear to be fiction – witness Apple's recent announcement not to offer AI technology ("Apple Intelligence") in the European Union for fear of violating the Digital Markets Regulation – Regulation (EU) 2022/1925 (https://www.theverge.com/2024/6/21/24183251/apple-eu-delay-ai-screen-mirroring-shareplay-dma) and Meta's announcement not to offer a more advanced model in view of the "too unpredictable nature" of the European regulatory environment (https://www.theverge.com/2024/7/18/24201041/meta-multimodal-llama-ai-model-launch-eu-regulations).

[136] 'La propuesta de Reglamento de AI: una intervencióin legislativa insoslayable en un contexto de incertidumbre' in MIGEL PEGUERA POCH (coord.), *Perspectivas Regulatios de La Inteligencia Artifical en La Unión Europea* (Reus 2023) p. 179.

[137] The Regulation itself acknowledges this in recital 121, which reads: "*Standardization should play a key role in providing providers with technical solutions that ensure compliance with this Regulation, in line with the state of the art, in order to promote innovation, competitiveness and growth in the single market*.". For a non-exhaustive list of *standards* applicable in this context *see* FEDERICO MARENGO, (n 77) pp. 196 ff. and ALESSIO TARTARO, 'Regulating by standards: current progress and main challenges in the standardization of Artificial Intelligence in support of the AI Act' European Journal of Privacy Law and Technologies (2023) pp. 147-174.